\documentclass[final,5p,times,twocolumn]{elsarticle}

\usepackage{graphicx}
\usepackage{amssymb}
\usepackage{amsmath}
\usepackage{color}

\journal{Physics Letters B}
\newcommand{\bea}{\begin{eqnarray}}
\newcommand{\eea}{\end{eqnarray}}
\newcommand{\be}{\begin{equation}}
\newcommand{\ee}{\end{equation}}
\newcommand{\bi}{\begin{itemize}}
\newcommand{\ei}{\end{itemize}}
\newcommand{\benu}{\begin{enumerate}}
\newcommand{\eenu}{\end{enumerate}}
\newcommand{\nn}{\nonumber}
\newfont{\bg}{cmr10 scaled\magstep4}

\begin{document}

\begin{frontmatter}

\title{CP-odd Higgs Boson Production in $e\gamma$ Collisions}

\author[yokohama]{Ken Sasaki}
\ead{sasaki@ynu.ac.jp}

\author[kyoto]{Tsuneo Uematsu}
\ead{uematsu@scphys.kyoto-u.ac.jp}

%--- institute information

\address[yokohama]{ Dept. of Physics, Faculty of Engineering\\
 Yokohama National University, Yokohama 240-8501, Japan}

\address[kyoto]{Institute for Liberal Arts and Sciences,
Kyoto University, Kyoto 606-8501, Japan\\
and Maskwa Institute, Kyoto Sangyo University, Kyoto 603-8555, Japan}

%--- abstract
\begin{abstract}
We investigate the CP-odd Higgs boson production via two-photon 
processes in $e\gamma$ collisions. The CP-odd Higgs boson, which we denote as 
$A^0$, is expected to appear in the Two-Higgs Doublet Models (2HDM) as
a minimal extension of Higgs sector for which the Minimal Supersymmetric 
Standard Model (MSSM) is a special case.
The scattering amplitude for $e\gamma\rightarrow eA^0$ is evaluated at the electroweak one-loop level. 
The dominant contribution  comes  
from top-quark loops when $A^0$ boson is rather light and $\tan\beta$ is not 
large. There are no contributions from the 
$W$-boson and $Z$-boson loops nor the scalar top-quark (stop) loops. 
The differential cross section for the $A^0$ production is analysed. 

\end{abstract}

%--- keywords
\begin{keyword}
CP-odd Higgs production, two-photon fusion, transition form factor, $e\gamma$-collisions
\end{keyword}

\end{frontmatter}

% \linenumbers

%\tableofcontents
%--------- main text --------------\
\section{Introduction \label{introduction}}
After the Higgs boson with mass about 125 GeV was discovered by ATLAS and CMS 
at LHC~\cite{HiggsLHC} and its spin, parity and couplings were 
examined~\cite{SpinParity}, 
there has been growing interest in constructing a new 
accelerator facility, like a linear $e^+e^-$ collider~\cite{ILC}, which would 
offer much cleaner experimental data. Along with $e^+e^-$ collider, other options such as $e^-e^-$, $e^-\gamma$ and $\gamma\gamma$ colliders have also been discussed. See Refs.~\cite{DeRoeck}-\cite{GK} and the references therein. 
Each option for colliders will provide interesting topics to study, such as the detailed measurement of the Higgs boson properties and the quest for the new 
physics beyond the Standard Model (SM). An $e^-e^-$ collider is easier 
to build than an $e^+e^-$ collider and may stand as a potential candidate 
before positron sources with high intensity are available. The $e^-\gamma$ 
and $\gamma\gamma$ options are based on $e^-e^-$ collisions, where one or 
two of the electron beams are converted to the photon beams.

In our previous papers~\cite{KWSUPL,WKUSPRD}, we have studied the SM Higgs 
boson ($ H_{\rm SM}$) production in $e\gamma$ collisions, focusing on the transition form factor of  $H_{\rm SM}$ boson~\cite{KWSUPL} and also on the dependence of polarizations of the initial electron and photon beams~\cite{WKUSPRD}. 
In this paper we investigate the production of the CP-odd Higgs boson ($A^0$), 
which appears in the 2HDM or in the MSSM~\cite{Hunter}, in an $e^- \gamma$ collider (Fig.1).
A originally proposed center of mass energy was 500 GeV for an $e^+e^-$ linear 
collider 
~\cite{ILC}. In the light of an $e^- \gamma$ collider,  we study for a case when $A^0$ boson is rather light. More specifically, we assume that the $A^0$ mass is less than 500 GeV.
We examine the reaction $e\gamma\rightarrow eA^0$ at the one-loop level in 
the electroweak interaction. Due to the absence of the tree-level $ZZA^0$ and $W^+W^-A^0$ couplings, the one-loop diagrams which contribute to the reaction are through the $\gamma^*\gamma$-fusion and $Z^*\gamma$-fusion processes. It 
turns out that the contribution of the $\gamma^*\gamma$-fusion diagrams is 
far more dominant over  the one from the $Z^*\gamma$-fusion diagrams. 
Thus the $A^0$ production in $e\gamma$ collisions is well-described
by the ``so-called" transition form factor~\cite{KWSUPL}. We investigate 
the $Q^2$ dependence of the transition form factor and  the production cross 
section. 

In the next section we briefly 
outline the CP-odd Higgs boson $A^0$
in the type-II 2HDM or in the MSSM. In section 3, we calculate the one-loop 
electroweak corrections to the $A^0$ production in $e\gamma$ collisions. 
We also discuss the transition form factor for the $\gamma^*\gamma$-fusion 
process in $e\gamma$ scattering. In section 4, we present the numerical 
analysis of the differential cross section for the $A^0$ production
and its dependence on the $A^0$ mass.
The final section is devoted to the concluding remarks. 

%%%%%%%%%%%%%%%%%%%%%%%%%%%%%%%%%%%%%
\begin{figure}[hbt]
\begin{center}
\includegraphics[scale=0.25]{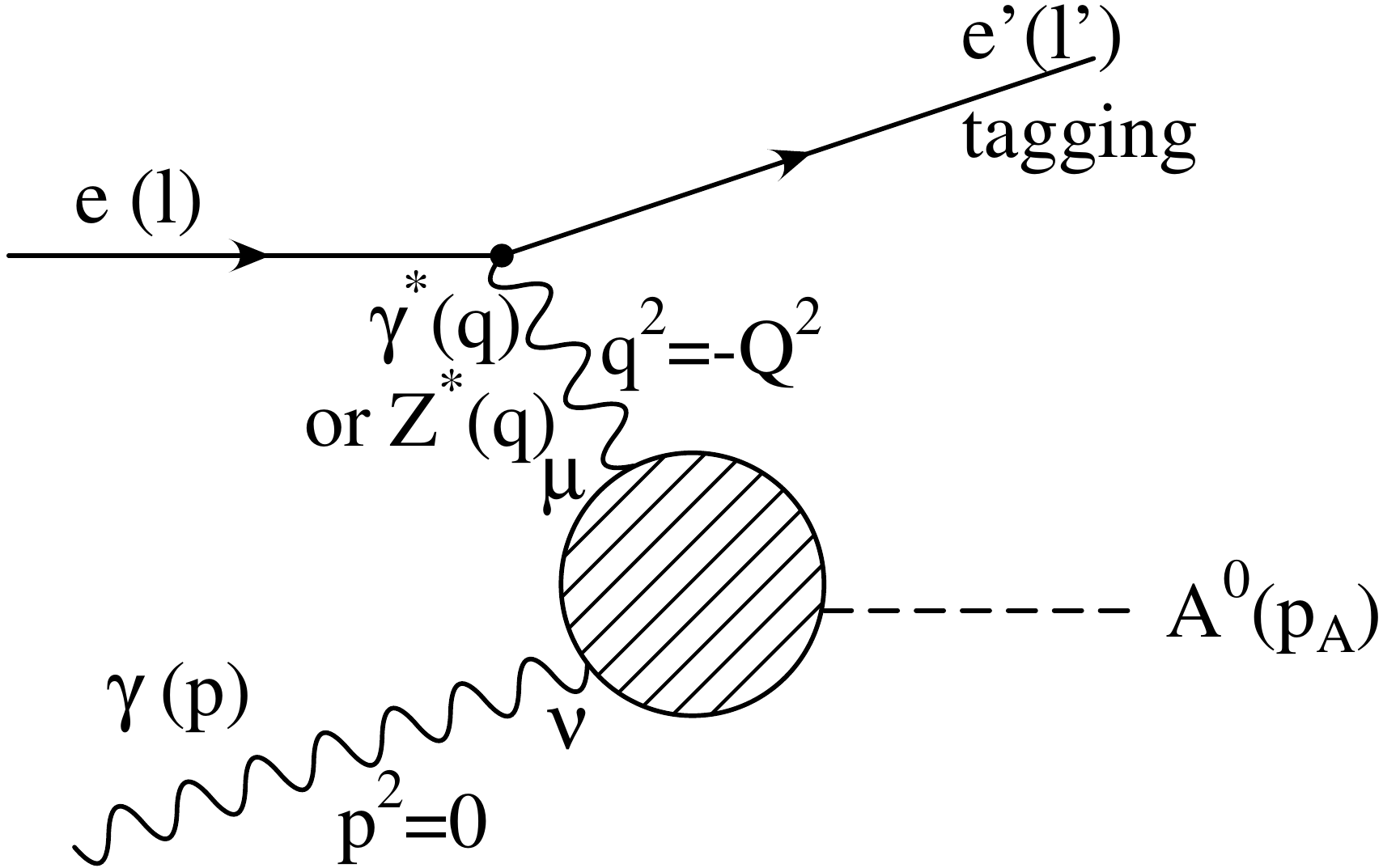}
%\vspace{-0.3cm}
\caption{\label{e-gamma-collision-A-blob} 
Production of CP-odd Higgs boson $A^0$ 
in the electron-$\gamma$ collision.
} 
\end{center}
\end{figure}

\vspace{-0.5cm}

%%%%%%%%%%%%%%%%%%%%%%%%%%%%%%%%%%%%%%%%%%%%%%%%%%%%%
\section{CP-odd Higgs boson in 2HDM/MSSM}\label{Section2}
%%%%%%%%%%%%%%%%%%%%%%%%%%%%%%%%%%%%%%%%%%%%%%%%%%%%%
As a minimal extension of the Higgs sector of the SM, we consider the type-II 2HDM which includes the MSSM as a special case~\cite{Hunter}.  
We denote the two SU(2)$_L$ doublets $H_1$ and $H_2$ with weak hypercharge
$Y=-1$ and $Y=1$, respectively, by the 
4 complex scalar fields, $\phi_1^0$, $\phi_1^-$, $\phi_2^+$, $\phi_2^0$
as follows:
\begin{eqnarray}
H_1=\left(
\begin{array}{c}
H_1^1\\
H_1^2
\end{array}
\right)=\left(
\begin{array}{c}
\phi_1^{0*}\\
-\phi_1^-
\end{array}
\right),
\quad
H_2=\left(
\begin{array}{c}
H_2^1\\
H_2^2
\end{array}
\right)=\left(
\begin{array}{c}
\phi_2^+\\
\phi_2^0
\end{array}
\right)~,
\end{eqnarray}
where, in the type-II model, $H_1$ ($H_2$)
couples only  to  down-type (up-type) quarks and leptons.
They acquire the following vacuum expectation values after the
spontaneous symmetry breaking:
\begin{eqnarray}
\langle H_1\rangle=\left(
\begin{array}{c}
v_1\\
0
\end{array}
\right),
\quad
\langle H_2\rangle=\left(
\begin{array}{c}
0\\
v_2
\end{array}
\right),\quad
\tan\beta=v_2/v_1~.
\end{eqnarray}
Then 3 degrees of freedom out of 8 consisting of the 4 complex scalar fields
are absorbed by the longitudinal components of $W^\pm$, $Z$,  
and the remaining 5 degrees of freedom become the following two charged and 
three neutral physical Higgs bosons:
\begin{eqnarray}
\mbox{Charged}\ \ H^+,\ H^-;\ \mbox{CP-even}\ \ h^0,\ H^0;\ \mbox{CP-odd}\ \ A^0~.
\end{eqnarray}
Here we are particularly interested in the CP-odd Higgs boson $A^0$ and 
investigate its production in $e\gamma$ collisions.
%%%%%%%%%%%%%%%%%%%%%%%%%%%%%

%%%%%%%%%%%%%%%%%%%%%%
We enumerate some characteristics of $A^0$ couplings to other fields in the 
type-II 2HDM and the MSSM.
\begin{itemize}
\item[1)]\
In contrast to the CP-even Higgs bosons 
$h^0$ and $H^0$, $A^0$ does not couple to 
$W^+W^-$ and $ZZ$ pairs at tree level. Hence $W$-boson and $Z$-boson one-loop 
diagrams do not contribute to the $A^0$ production.
\item[2)]\
$A^0$ does not couple to other two physical Higgs bosons in  cubic interactions.
\item[3)]\
The couplings of $A^0$ to quarks and leptons are proportional to their masses. 
Therefore, we  consider only the top and bottom quark-loop diagrams for the $A_0$ production. 
The $A^0$ coupling to the top (bottom) with mass $m_t$ ($m_b$) is given by $\lambda_t \gamma_5$ ($\lambda_b \gamma_5$) with ~\cite{Hunter}
\bea
\lambda_t &=&-\frac{gm_t\cot\beta}{2m_W}\equiv g{\widetilde \lambda}_t m_t,
\label{lambdatop}\\
\lambda_b &=&-\frac{gm_b\tan\beta}{2m_W}\equiv g{\widetilde \lambda}_b m_b.
\label{lambdabottom}
\eea
Here $g$ and $m_W$ are the weak gauge coupling and the weak boson mass, 
respectively. 

\quad In the MSSM, charginos also couple to $A^0$.  When $CP$ is conserved (which we assume in this paper), the diagonal couplings of $A^0$ to the chargino mass eigenstates are purely pseudoscalar~\cite{GunionHaber}, whose couplings are expressed as $g\kappa_i \gamma_5$ with (see Eq.(4.32) of ~\cite{GunionHaber}),
\be
\kappa_i=\frac{1}{\sqrt{2}}\Bigl( \sin\beta U_{i2} V_{i1} +\cos\beta U_{i1} V_{i2} \Bigr),\qquad 
i=1,2~,
\ee
where $U$ and $V$ are $2\times 2$ orthogonal matrices.  Thus $\kappa_i \sim {\cal O}(1)$. 
In the following we deal with two chargino mass eigenstates as a whole and write its coupling to 
$A_0$ and mass as $\kappa$ and $m_\chi$, respectively. We put
\be
\lambda_\chi=g\kappa\equiv g{\widetilde \lambda}_\chi m_\chi~. \label{lambdachi}
\ee
Recently at LHC, ATLAS~\cite{ATLASchargino} and CMS~\cite{CMSchargino} excluded chargino masses below 1140 GeV for the case that the lightest supersymmetric particles are massless
~\cite{RPPexperiment}. 
The results depend on the various scenarios for the production and decay of charginos and neutralinos. We therefore take $m_\chi=1$ TeV as a benchmark mass for chargino in this paper.

\item[4)]\ In the case of the MSSM, the trilinear $A^0$ coupling
to mass-eigenstate squark pairs $\tilde{q}_i\tilde{q}_i$ ($i=1,2$) vanishes
\cite{Hunter}. 
Hence, the scalar top-quark (stop) does not contribute to the $A^0$ production 
in $e\gamma$ collisions at one-loop level.
\end{itemize}

%%%%%%%%%%%%%%%%%%%%%%%%%%%%%%%%%%%%%%%%%%%%%%%%%%%%%%%%%%%%%%
\section{CP-odd Higgs Boson Production in $e\gamma$ Collisions}
\begin{figure}[hbt]
\begin{center}
\includegraphics[scale=0.25]{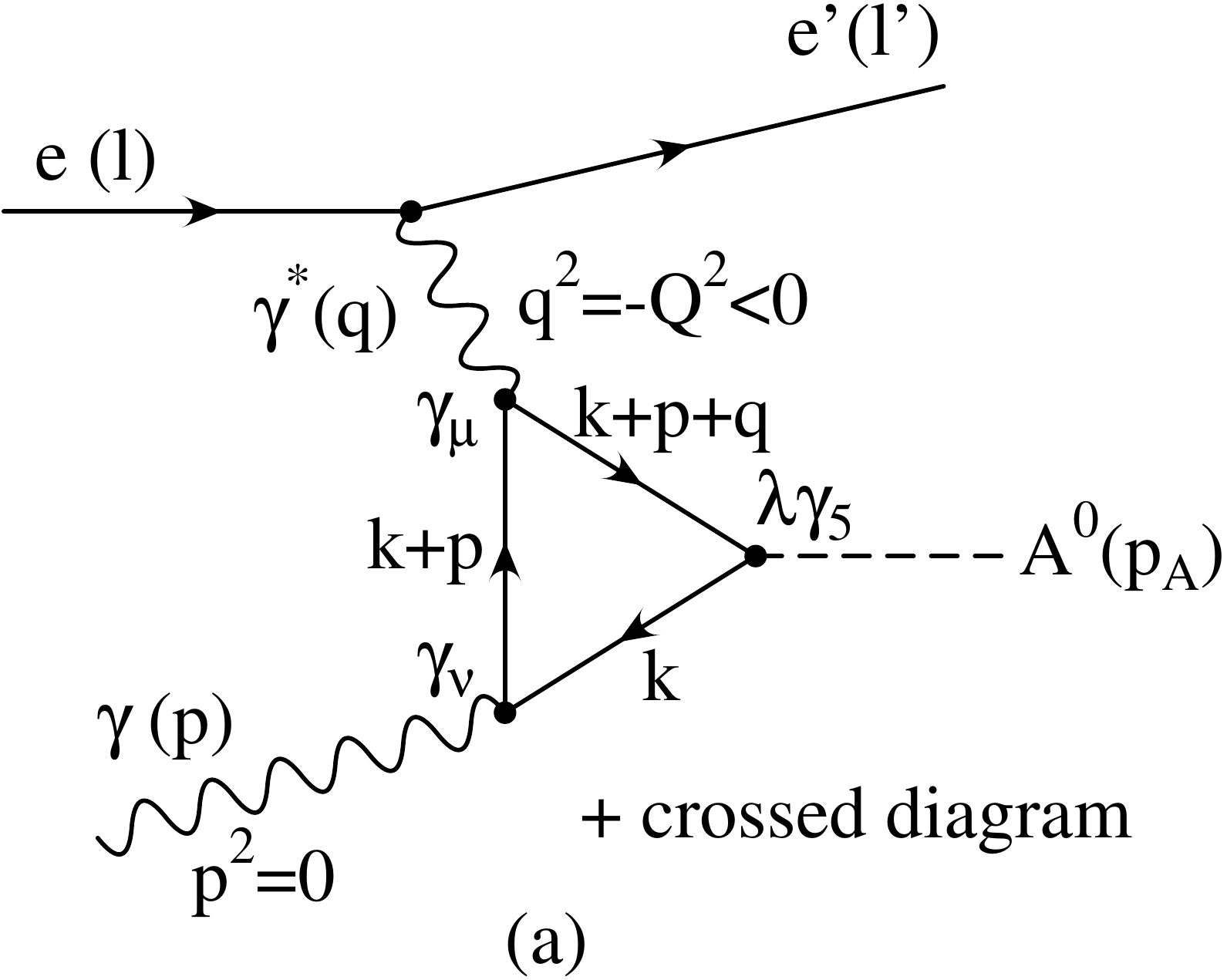}
\
\includegraphics[scale=0.25]{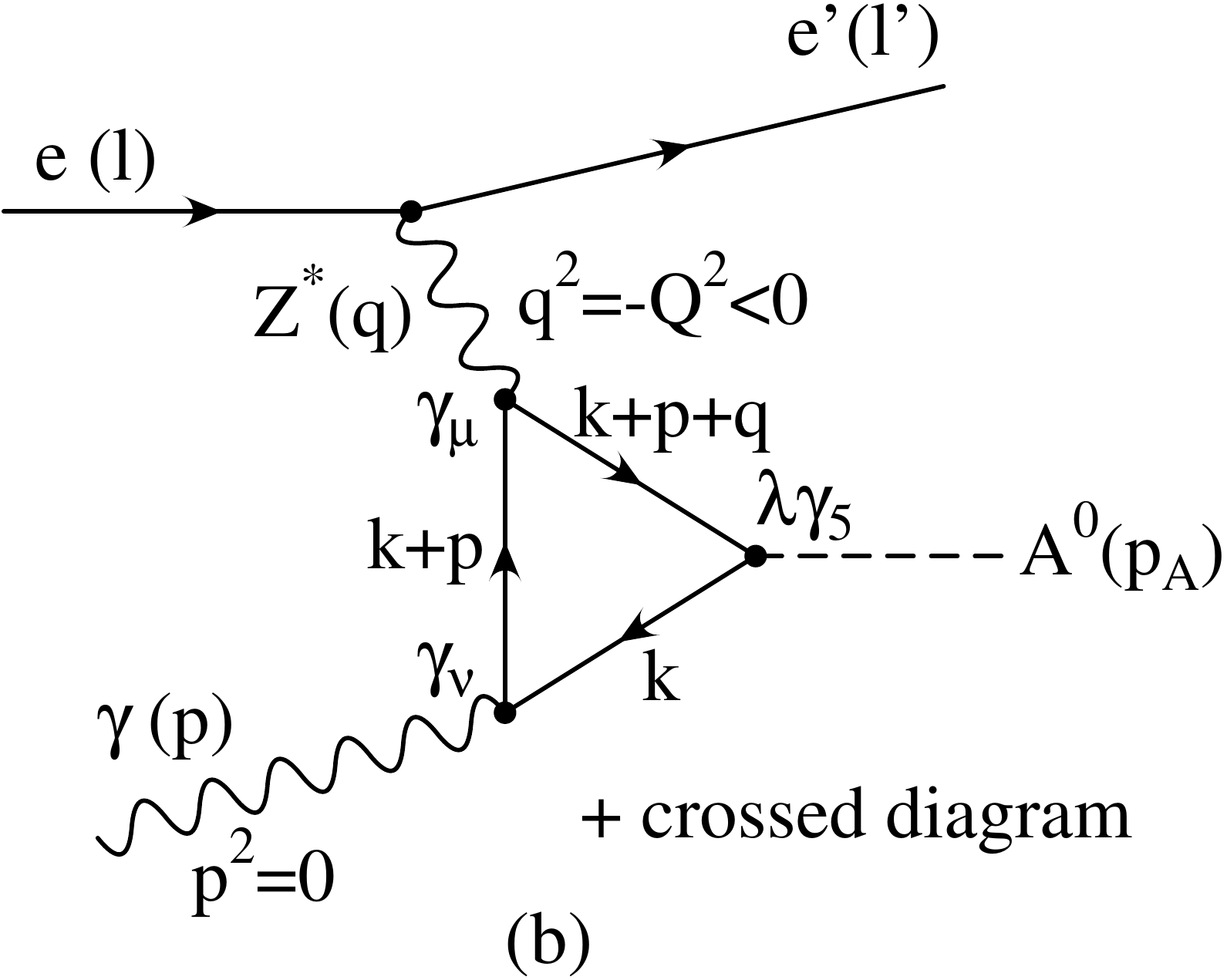}
\caption{\label{ggandZgfusion} 
(a) $\gamma^*\gamma$ fusion diagrams for $e \gamma\rightarrow e'A^0$. 
\ (b) $Z^*\gamma$ fusion diagrams for $e \gamma\rightarrow e'A^0$.} 
\end{center}
\end{figure}

%%%%%%%%%%%%%%%%%%%%%%%%%%%%%%%%%%%%%%%%%%%%%%%%%%%%%%%%%%%%%%
We 
investigate the production of the CP-odd Higgs boson $A^0$ in an 
$e\gamma$ collision experiment  (Fig.\ref{e-gamma-collision-A-blob}):
\bea
  e(l) +\gamma(p) \rightarrow 
 e(l') +A^0(p_A)~, \label{CP-oddHiggsProduction}
\eea
where we detect the scattered electron in the final state.
The one-loop diagrams which contribute to the reaction
 (\ref{CP-oddHiggsProduction}) 
are classified into two groups:  $\gamma^*\gamma$ fusion diagrams and $Z^*\gamma$ fusion diagrams (Fig.\ref{ggandZgfusion}).
As we will see later, the contribution of the 
former is far more dominant over that of the latter.

%%%%%%%%%%%%%%%%%%%%%%%%%%%
%%%%%%%%%%%%%%%%%%%%%%%%%%%
Since
$p$ is the momentum of a real photon,  we have $p^2=0$.  We set 
$q=l-l'$. Assuming that electrons are massless so that $l^2={l'}^2=0$, 
we introduce the following Mandelstam variables:
\bea
&&s=(l+p)^2=2l\cdot p,\quad t=(l-l')^2=q^2\equiv-Q^2,\\
&&u=(l-p_A)^2=m_A^2-s-t~.
\eea
where $p_A^2=m_A^2$ with $m_A$ being the 
$A^0$ boson mass.

\subsection{One-loop $\gamma^*\gamma$ fusion diagrams}

Due to the characteristics of $A^0$ couplings to other fields, we take into account  only 
the loops of three fermions (top ($t$) and bottom ($b$) quarks and chargino ($\chi$))  for the $\gamma^*\gamma$ fusion diagrams (Fig.\ref{ggandZgfusion} (a)). 
The contribution from the one-loop $\gamma^*\gamma$ fusion diagrams to the scattering amplitude is expressed  as
\bea
\langle e' A^0|T|e\gamma\rangle_{\gamma^*\gamma}^f=
[\overline{u}(l')(-ie\gamma_\mu)u(l)]\frac{-i}{q^2+i\epsilon}A_{\mu\nu}^f\epsilon^\nu(p) ,\label{Contributiongammagamma}
\eea
where $u(l)$ 
($\overline{u}(l')$)is the spinor for the initial (scattered) electron
with momentum $l$ ($l')$ and $\epsilon^\nu(p)$ is the photon polarization vector with $p_\nu
\epsilon^\nu(p)=0$.
The tensor $A^f_{\mu\nu}$ with $f=t,b,\chi$ is given as
\bea
&&\hspace{-1.3cm}
A^f_{\mu\nu}=8N^f_C q_f^2e^2\lambda_f m_f\varepsilon_{\mu\nu\alpha\beta}q^\alpha 
p^\beta
 \frac{1}{16\pi^2}C_0(0,q^2,m_A^2; m_f^2,m_f^2,m_f^2),\nonumber\\
\label{Amunu}
\eea
where $e$ is the electromagnetic coupling, $N_C^f$ is a color factor with $N^t_C=N_C^b=3$,  $N^\chi_C=1$, $q_f$ is a charge factor with $q_t=\frac{2}{3}$, $q_b=-\frac{1}{3}$, $q_\chi=1$ and $C_0$ is a Passarino-Veltman three-point scalar integral~\cite{PassarinoVeltman}:
\bea
&&\hspace{-1.3cm}C_0(p^2,q^2,(p+q)^2; m_f^2,m_f^2,m_f^2)\nonumber \\
&&\hspace{-1.3cm}=\frac{1}{i\pi^2}\int 
\frac{d^4k}{[k^2-m_f^2][(k+p)^2-m_f^2][(k+p+q)^2-m_f^2]}~.
\eea
The integral $C_0$ is expressed as the sum of two functions $f(\tau_f)$ and $g(\rho_f)$ as
\bea
\hspace{-0.7cm}C_0(0, -Q^2,m_A^2; m_f^2,m_f^2,m_f^2)
=-\frac{1}{Q^2+m_A^2}\left\{2f(\tau_f)+\frac{1}{2}g(\rho_f)\right\}, \label{C0}
\eea
where the dimensionless variables $\tau_f$ and $\rho_f$ are defined as
\bea
\tau_f\equiv \frac{4m_f^2}{m_A^2}, \quad \rho_f\equiv \frac{Q^2}{4m_f^2}~,
\eea
and 
\bea
f(\tau)&=&\left[\sin^{-1}\sqrt{\frac{1}{\tau}}\right]^2 \hspace{2.8cm} 
\tau\geq 1 \label{ftau-1}~,\\
&=&-\frac{1}{4}\left[\log{\frac{1+\sqrt{1-\tau}}{1-\sqrt{1-\tau}}}-i\pi
\right]^2 \qquad \tau<1\label{ftau-2}~,\\
g(\rho)&=&\left[\log\frac{\sqrt{\rho+1}+\sqrt{\rho}}
{\sqrt{\rho+1}-\sqrt{\rho}}\right]^2~.
\eea
Similar combinations of functions $f(\tau)$ and $g(\rho)$ as in Eq.(\ref{C0}) with the time-like virtual mass, 
which are different from our space-like case, appear in the Higgs decay processes $H_{\rm SM}\rightarrow \gamma^* \gamma$ and $H_{\rm SM}\rightarrow Z^* \gamma$  in Ref.\cite{RomaoAndringa1997} (see also Ref.\cite{Hunter} for on-shell decays, 
$H_{\rm SM}\rightarrow \gamma \gamma$~\cite{H2gammas} and $H_{\rm SM}\rightarrow Z \gamma$).

%%%%%%%%%%%%%%%%%%%%%%%%%%%%%%%%%%%%
\subsection{Transition Form Factor}
%%%%%%%%%%%%%%%%%%%%%%%%%%%%%%%%%%%%
Inserting the expressions of $\lambda_f$ given in Eqs.(\ref{lambdatop}), (\ref{lambdabottom}) and  
(\ref{lambdachi}) back to Eq.(\ref{Amunu}), we 
see that $A^f_{\mu\nu}$ is expressed as
\bea
A^f_{\mu\nu}=-\frac{e^2g}{(4\pi)^2} N^f_C q_f^2 {\widetilde \lambda}_fF_f(Q^2,m_A^2, m_f^2)\ \varepsilon_{\mu\nu\alpha\beta}q^\alpha p^\beta~,
\eea
where we have introduced a transition form factor given by
\bea
F_f(Q^2,m_A^2, m_f^2)&=&\frac{\tau_f}{1+\rho_f\tau_f}[g(\rho_f)+4f(\tau_f)]\nn\\
&&\hspace{-0.7cm}=-8m_f^2C_0(0, -Q^2,m_A^2; m_f^2,m_f^2,m_f^2).\label{FormFactor}
\eea
%A dimensionless quantity $N_Cq_t^2 F(Q^2,m_A^2, m_t^2)$ may be considered as a transition from factor of the $A^0$ boson. 
Note that
for $m_A<2m_f$, \ $i.e.$\  $\tau_f>1$,  $f(\tau_f)$ is given by Eq.(\ref{ftau-1})
which is a real function, while for $m_A>2m_f$, \ $i.e.$\ $\tau_f<1$ we have $f(\tau_f)$ given by Eq.(\ref{ftau-2}) which is a complex function.

\begin{figure}[hbt]
\begin{center}
\includegraphics[scale=0.3]{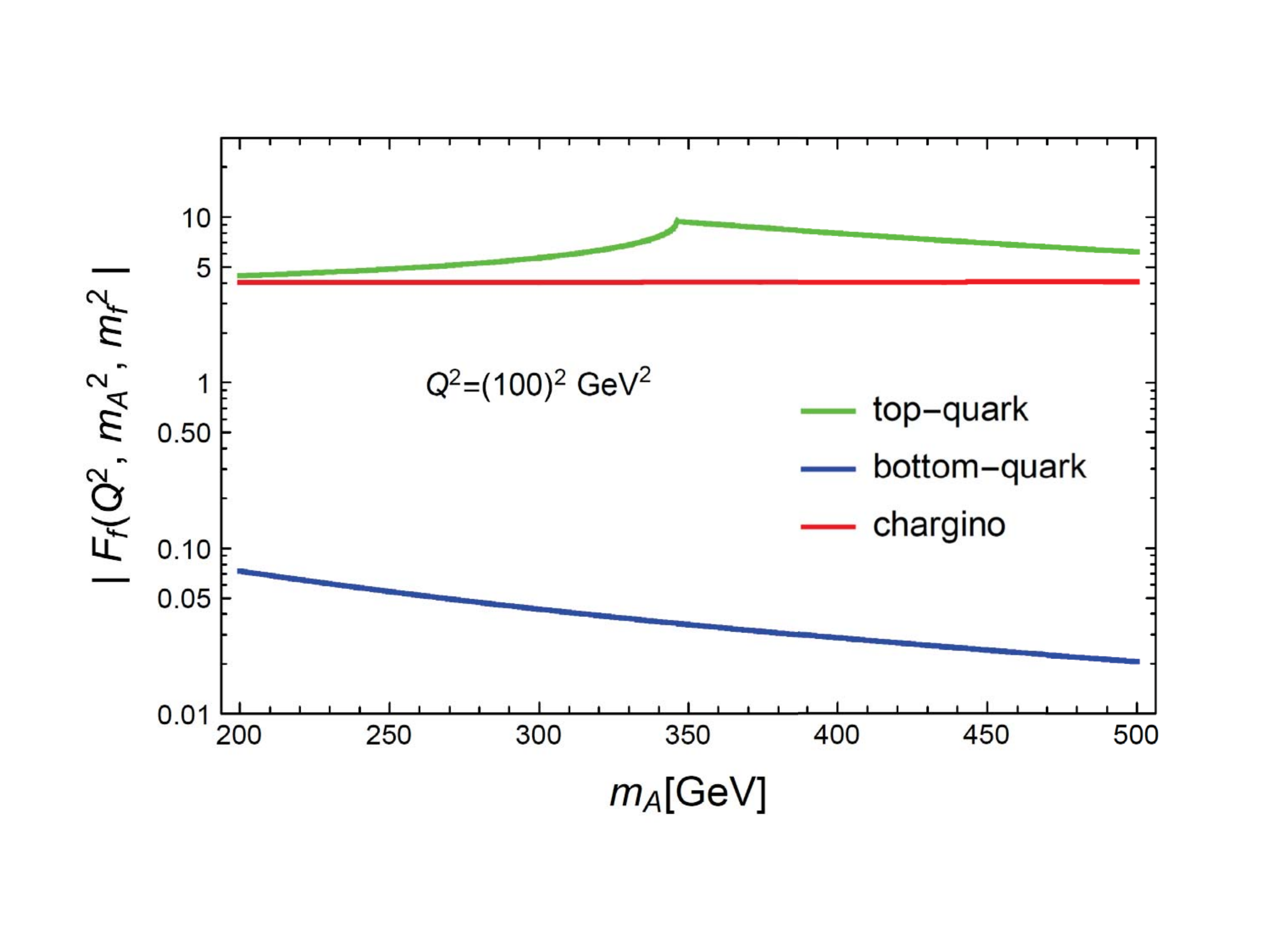}
\caption{$|F_f(Q^2, m_A^2, m_f^2)|$ as a function of $m_A$ with $Q^2=100^2 {\rm GeV}^2$: 
top-quark (green line),  bottom-quark (blue line) and  chargino with mass $m_\chi=1000{\rm GeV}$ (red line)\label{mAQ100}. }
\end{center}
\end{figure}

Taking the mass parameters as $m_t=173 {\rm GeV}$, $m_b=4.3 {\rm GeV}$ and 
 $m_{\chi}= 1000 {\rm GeV}$, we analyze     
the behaviours of $|F_f(Q^2, m_A^2, m_f^2)|$.  We plot $|F_f(Q^2, m_A^2, m_f^2)|$ in Fig.\ref{mAQ100} as a function of $m_A$ for the case $Q^2=(100)^2 {\rm GeV}^2$. 
Note that $|F_f(Q^2, m_A^2, m_f^2)| \rightarrow 4$ as $m_f \rightarrow \infty$, while 
$|F_f(Q^2, m_A^2, m_f^2)| \rightarrow 0$ as $m_f \rightarrow 0$. 
We see a kink structure at the threshold region $m_A\approx 2m_t$ for $|F_t(Q^2, m_A^2, m_f^2)|$. 
Fig.\ref{mAQ100} shows that the ratio $|F_b(Q^2, m_A^2, m_b^2)|/|F_t(Q^2, m_A^2, m_t^2)|\sim 0.003-0.016$ and $|F_\chi(Q^2, m_A^2, m_\chi^2)|$ is the same order as 
$|F_t(Q^2, m_A^2, m_t^2)|$ when $m_\chi = 1000 {\rm GeV}$

On the other hand, we obtain 
\bea
|N_C^bq_b^2{\widetilde \lambda}_b|/|N_C^tq_t^2{\widetilde \lambda}_t|
&=&\frac{\tan^2\beta}{4}~, \\
|N_C^\chi q_\chi^2{\widetilde \lambda}_\chi|/|N_C^tq_t^2{\widetilde \lambda}_t|
&=&\frac{3}{2}\frac{m_W}{m_\chi}|\kappa| \tan\beta~. 
\eea 
Direct searches for heavy neutral Higgs bosons have been performed at  LHC. The results were  interpreted in the MSSM benchmark scenarios. In the context of the hMSSM 
scenario~\cite{hMSSM}, ATLAS data~\cite{ATLAStanbeta} excluded $\tan\beta>1.0$ for $m_A=250$GeV and $\tan\beta>42.0$ for $m_A=1.5$TeV at the 95\% CL.
Here in this paper we are dealing with a rather light $A^0$ boson with mass $m_A\le 500$GeV. Therefore we consider the case where $\tan\beta$ is not large, e.g. $\tan^2\beta \leq 10$.

%%%%%%%%%%%%%%%%%%%%%%%%%%%%%%%%%
The production cross section is proportional to the absolute square of the amplitude.
Hence the ratio of the bottom-quark (charginos) contribution to the 
one of top-quark is given as the square of the quantity in Eq.(21) (Eq.(22)) 
multiplied by $|F_b|^2/[F_t|^2$ ($|F_\chi|^2/[F_t|^2$).
Then we find that for the case $\tan^2\beta \leq 10$ we can ignore the contributions from
the bottom-quark and charginos as compared to the one from top-quark. When $\tan\beta \simeq 10$ we can still
neglect the bottom-quark contribution but the chargino's contribution 
becomes the same order as the top-quark contribution.
%%%%%%%%%%%%%%%%%%%%%%%%%%%%%%%%%

In the following we proceed with our analysis of the reaction $e\gamma\rightarrow eA^0$ 
assuming that  $A^0$ boson is rather light and $\tan\beta$ is not large. 

%%%%%%%%%%%%%%%%%%%%%%%%%%%%%%%%%%%%%%%%%%%%%%%%%%%%%%%%%%%%%%%
\subsection{One-loop $Z^*\gamma$ fusion diagrams}
%\subsection{Z boson and real $\gamma$ fusion contribution}
%%%%%%%%%%%%%%%%%%%%%%%%%%%%%%%%%%%%%%%%%%%%%%%%%%%%%%%%%%%%%%%%%

The one-loop $Z^*\gamma$ fusion diagrams for the $A^0$ production are obtained from the one-loop 
$\gamma^*\gamma$ fusion diagrams by replacing the photon propagator with that of the $Z$ boson with mass $m_Z$ (Fig.\ref{ggandZgfusion} (b)).  
The loop contributions from three fermions (top ($t$) and bottom ($b$) quarks and chargino ($\chi$)) are expressed in terms of the function $F_f(Q^2,m_A^2, m_f^2)$ in Eq.(\ref{FormFactor}). Since the coupling strengths of $Z\cdot t\cdot t$, $Z\cdot b\cdot b$ and $Z\cdot \chi\cdot \chi$ vertices are the same order of magnitude, the argument in the previous subsection again follows: 
we can ignore the contributions from the bottom-quark and charginos for the case when $A^0$ boson is rather light and $\tan\beta$ is not large while the chargino mass is around 1TeV.

We consider the top quark loop contribution to the $Z^*\gamma$ fusion diagrams and obtain
\bea
&&\hspace{-1.3cm}\langle e' A^0|T|e\gamma\rangle^t_{Z^*\gamma}\nonumber\\
&&\hspace{-1.3cm}=\frac{g}{4\cos\theta_W}[\overline{u}(l')
(i\gamma_\mu)(f_{Ze}+\gamma_5)u(l)]\frac{-i}{q^2-m_Z^2}\widetilde{A}^t_{\mu\nu}
\epsilon^\nu(p)~, \label{ContributionZgamma}
\eea
with
\bea
\widetilde{A}^t_{\mu\nu}&=&8 N^t_Cq_t e\frac{g}{4\cos\theta_W}m_t\lambda_t f_{Zt}\epsilon_{\mu\nu\alpha\beta}q^\alpha p^\beta \nonumber\\
&&\times\frac{1}{16\pi^2}
C_0(0, -Q^2,m_A^2; m_t^2,m_t^2,m_t^2)~,
\eea
where $f_{Ze}$ and $f_{Zt}$ are the strength of vector part of the $Z$-boson coupling to 
electron and top quark, respectively, and are given by
\bea
f_{Ze}=-1+4\sin^2\theta_W~,\qquad f_{Zt}=1-\frac{8}{3}\sin^2\theta_W~,\label{fZefZt}
\eea
with $\theta_W$ being the Weinberg angle. In terms of the function  $F_t$ given in Eq.(\ref{FormFactor}),  
$\widetilde{A}^t_{\mu\nu}$ is rewritten as 
\bea
\widetilde{A}^t_{\mu\nu}=-\frac{eg^2 N^t_Cq_t {\widetilde \lambda}_t f_{Zt}}{(4\pi)^24\cos\theta_W}F_t(Q^2,m_A^2, m_t^2)\ \varepsilon_{\mu\nu\alpha\beta}q^\alpha p^\beta
\eea

%%%%%%%%%%%%%%%%%%%%%%%%%%%%%%%%%%%%%%%%%%%%%%%%%%%%%%%%%%%%%%%%%
\subsection{Differential cross section}
%%%%%%%%%%%%%%%%%%%%%%%%%%%%%%%%%%%%%%%%%%%%%%%%%%%%%%%%%%%%%%%%%
Adding two amplitudes $\langle e' A^0|T|e\gamma\rangle^t_{\gamma^*\gamma}$ and $\langle e' A^0|T|e\gamma\rangle^t_{Z^*\gamma}$ given in Eqs.(\ref{Contributiongammagamma}) and (\ref{ContributionZgamma}), we calculate
the differential cross section for the $A^0$ production   
in $e\gamma$ collisions with unpolarized initial beams, which turns out to be the sum of three terms:
\bea
&&\hspace{-1.5cm}\frac{d\sigma_{(\gamma^*\gamma)}}{dt}=\frac{\alpha_{\rm em}^3}{64\pi}
\frac{g^2}{4\pi}\Bigl(\frac{\cot\beta}{2m_W}\Bigr)^2\frac{1}{-t}\Bigl[1+\frac{u^2}{s^2}\Bigr]
\Bigl|N^t_Cq_t^2F_t(Q^2,m_A^2, m_t^2)\Bigr|^2~,\label{gamma-gamma-cross}\nn\\
&&\\
&&\hspace{-1.5cm}\frac{d\sigma_{(Z^*\gamma)}}{dt}=\frac{\alpha_{\rm em}}{64\pi}
\Bigl(\frac{g^2}{4\pi}\Bigr)^3\Bigl(\frac{\cot\beta}{2m_W}\Bigr)^2\Bigl( \frac{1}{16\cos^2\theta_W} \Bigr)^2\frac{-t}{(t-m_Z^2)^2}\Bigl[1+\frac{u^2}{s^2}\Bigr]
\nn\\
&&  \times f_{Zt}^2(f_{Ze}^2+1) \Bigl|N^t_Cq_tF_t(Q^2,m_A^2, m_t^2)\Bigr|^2~,\label{Z-gamma-cross}\\
&&\hspace{-1.5cm}\frac{d\sigma_{(\rm interference)}}{dt}=-2\times \frac{\alpha^2_{\rm em}}{64\pi}
\Bigl(\frac{g^2}{4\pi}\Bigr)^2\Bigl(\frac{\cot\beta}{2m_W}\Bigr)^2 \frac{1}{16\cos^2\theta_W} \frac{-1}{t-m_Z^2}\nn\\
&&\times \Bigl[1+\frac{u^2}{s^2}\Bigr] f_{Zt} f_{Ze}q_t \Bigl|N^t_Cq_tF_t(Q^2,m_A^2, m_t^2)\Bigr|^2~,\label{interference-cross}
\eea
where each corresponds to the contribution of the $\gamma^* \gamma$ fusion diagrams, 
the  $Z^* \gamma$ fusion diagrams and their interference, respectively, and
$\alpha_{\rm em}=e^2/4\pi$. 

%%%%%%%%%%%%%%%%%%%%%%%%%%%%%%%%%%%%%
\section{Numerical analysis}
%%%%%%%%%%%%%%%%%%%%%%%%%%%%%%%%%%%%%%
We analyze numerically the three differential cross sections given in Eqs.(\ref{gamma-gamma-cross})-(\ref{interference-cross}). We choose the mass parameters and the coupling constants as follows:
\bea
&&\hspace{-1cm} m_t=173~{\rm GeV}~,\quad m_Z=91~{\rm GeV}~,\quad m_W=80~{\rm GeV}~,\nn\\
&&\hspace{-1cm} \cos\theta_W=\frac{m_W}{m_Z}~,\quad e^2=4\pi\alpha_{em}=\frac{4\pi}{128}~,\quad 
g=\frac{e}{\sin\theta_W}~.\label{Nume}
\eea
The electromagnetic constant $e^2$ is chosen to be the value at the scale of 
$m_Z$. 
From Eqs.(\ref{fZefZt}) and (\ref{Nume}), we find $ f_{Zt} f_{Ze}<0$ and, therefore,  
Eq.(\ref{interference-cross}) shows that the interference between the $\gamma^* \gamma$ and $Z^* \gamma$ fusion diagrams works constructively and $\displaystyle{\frac{d\sigma_{({\rm Interference})}}{dt}}$ is positive.

\begin{figure}[hbt]
\begin{center}
\includegraphics[scale=0.37]{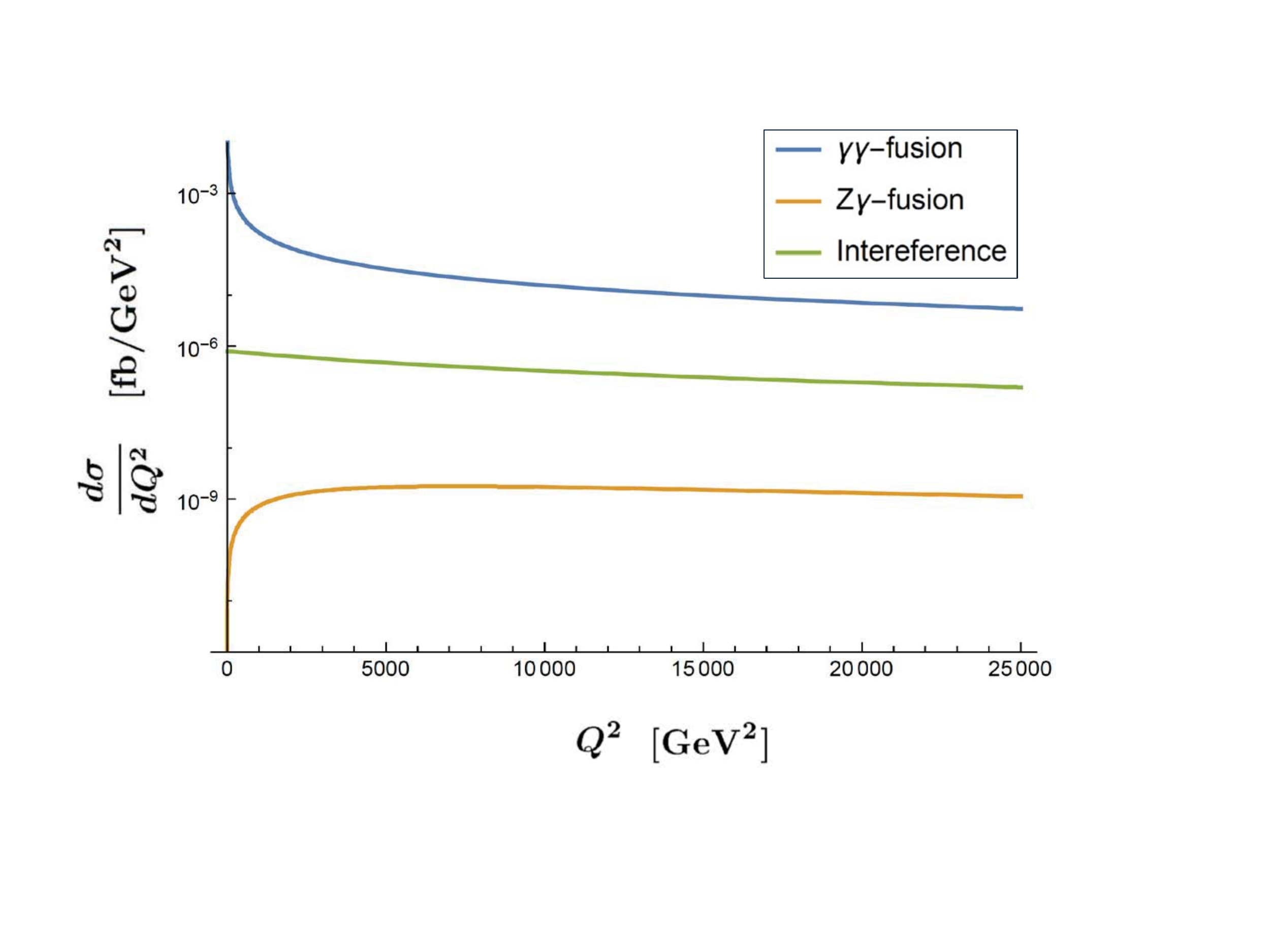}
\vspace{-1.5cm}
\caption{Comparison of the contribution among three differential cross sections for $\sqrt{s}=500$GeV, $m_A=400$GeV and $\cot\beta=1$.}
\label{3-processes} 
\end{center}
\end{figure}

We plot these differential cross sections as a function of $Q^2$ in Fig.\ref{3-processes} for the case
$\sqrt{s}=500$GeV, $m_A=400$GeV and $\cot\beta=1$.  (In fact,  the cross sections are proportional to 
$\cot^2\beta$.) 
We find that the contribution from the $\gamma^*\gamma$ fusion diagrams is far
more dominant over those from $Z^*\gamma$-fusion diagrams as well as from the interference 
term.
%%%%%%%%%%%%%%%%%%%%%
%%%%%%%%%%%%%%%%%%%
Actually we observe that at $Q^2=1000\ (5000)$ GeV$^2$,  the ratio of
$\displaystyle{\frac{d\sigma_{(Z^*\gamma)}}{dQ^2}}$ to 
$\displaystyle{\frac{d\sigma_{(\gamma^*\gamma)}}{dQ^2}}$ is 
$4.3\times 10^{-6}\ (5.2\times 10^{-5})$ 
and $\displaystyle{\frac{d\sigma_{({\rm Interference})}}{dQ^2}}$ to 
$\displaystyle{\frac{d\sigma_{(\gamma^*\gamma)}}{dQ^2}}$ is $4.1\times 10^{-3}\ (1.4\times 10^{-2})$. 
Thus the $A^0$ production in $e\gamma$ collisions is well-described by the 
$\gamma^*\gamma$ fusion diagrams with the top quark loop. 
This means that the transition form factor of the $A^0$ boson defined as $N^t_Cq_t^2 F_t(Q^2,m_A^2, m_t^2)$ 
in Eq.(\ref{FormFactor}) indeed makes sense and may be measurable in  $e\gamma$ collider experiments.  

\begin{figure}[hbt]
\vspace*{-0.8cm}
\begin{center}
\hspace*{-1.2cm}
\includegraphics[scale=0.37]{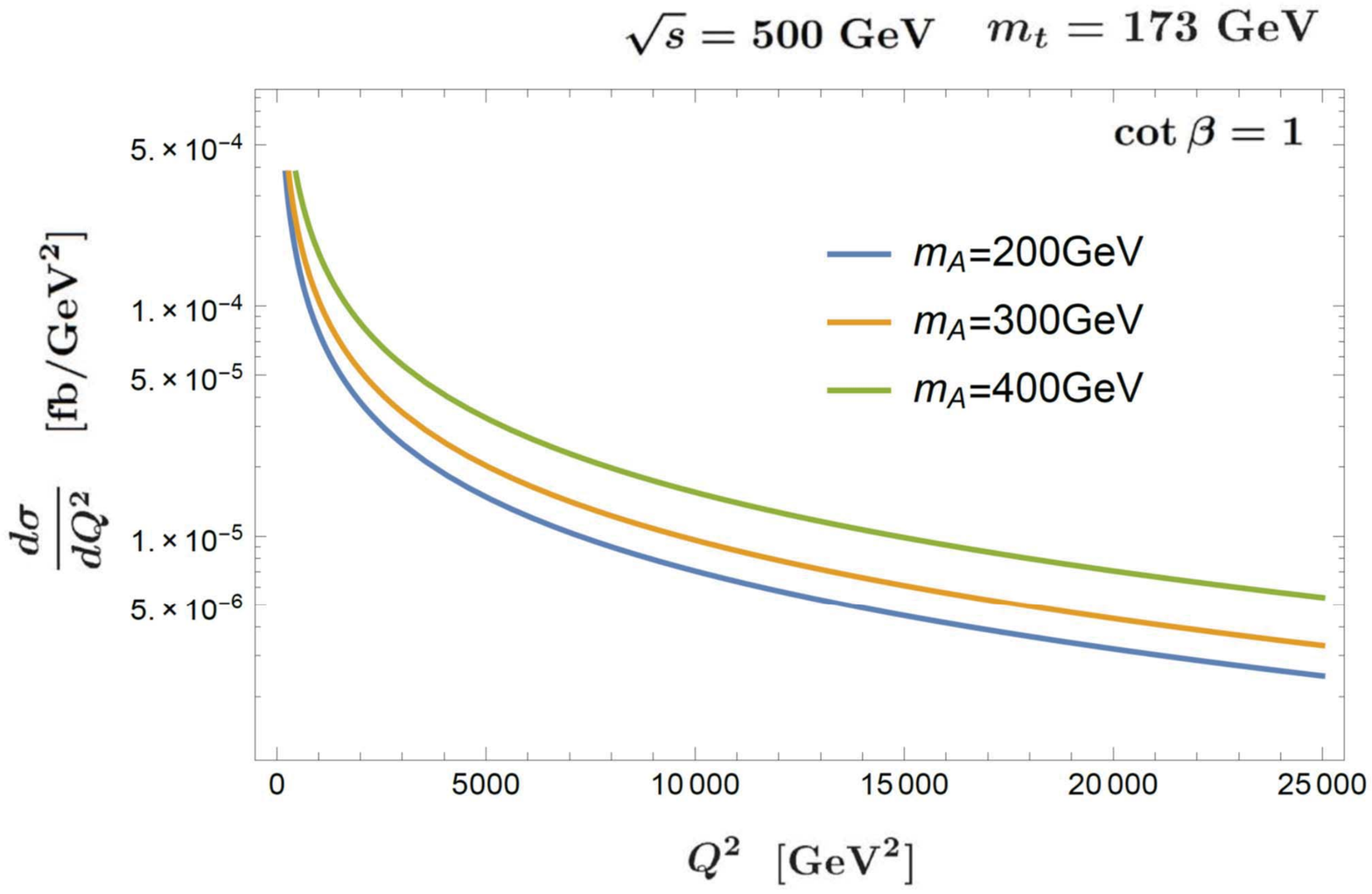}
\caption{Differential cross section for the production of CP-odd Higgs boson 
$A^0$ with mass $m_A=200,\ 300,\ 400$ GeV.}
\label{two-photon-process-3mass} 
\end{center}
\end{figure}

\begin{figure}[hbt]
\vspace*{-0.8cm}
\begin{center}
\hspace*{-1.2cm}
\includegraphics[scale=0.37]{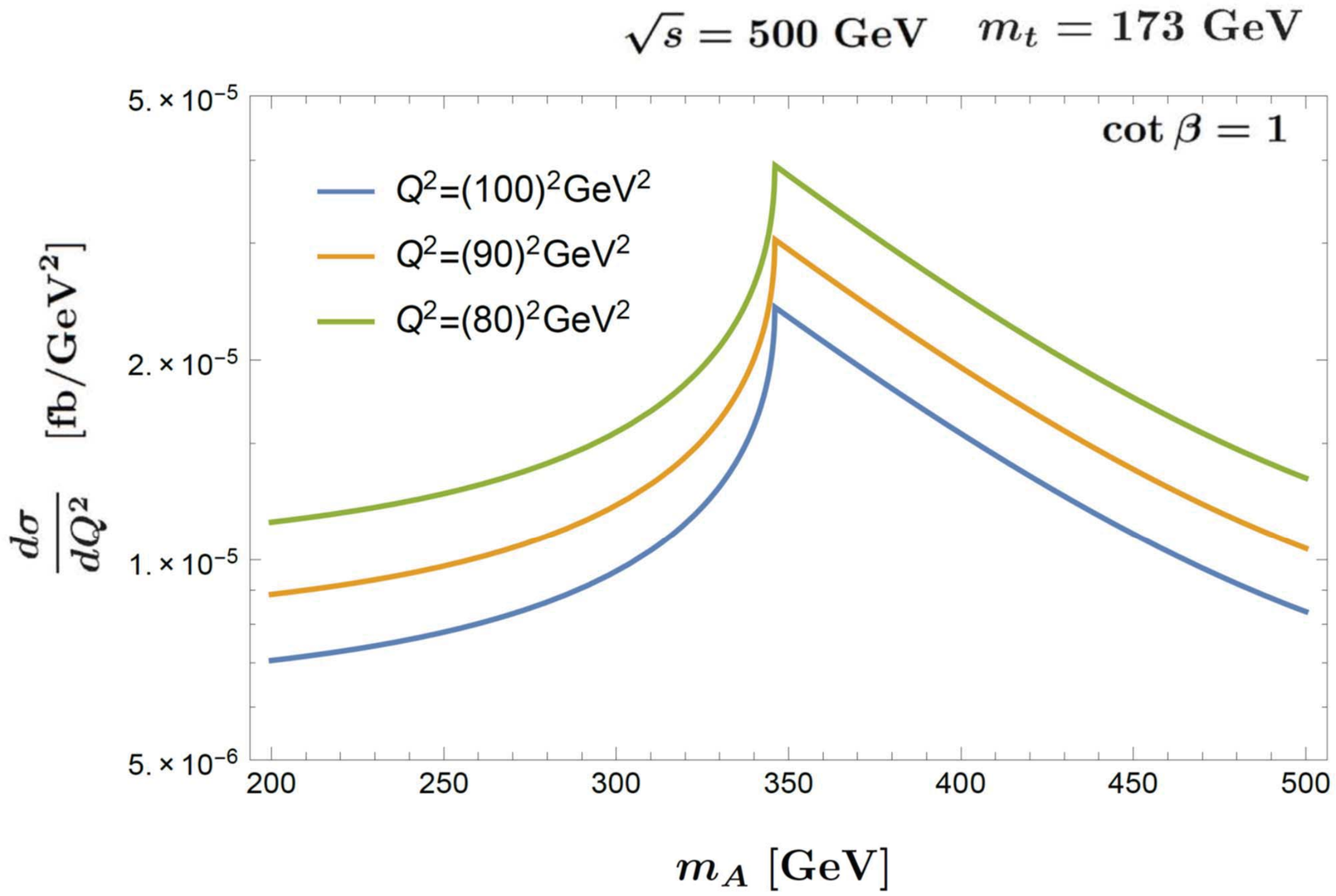}
\caption{The $A^0$ mass dependence of the differential cross section with $Q^2=(80)^2$GeV$^2$, $(90)^2$GeV$^2$,$(100)^2$GeV$^2$}
\label{mass-diff-dependence} 
\end{center}
\end{figure}

Now we shall focus on the $\gamma^*\gamma$ fusion process based on the 
formula for the production cross section given in Eq.(\ref{gamma-gamma-cross}).
In Fig.\ref{two-photon-process-3mass} we plot the differential
production cross section of $A^0$ with mass $m_A=200,\ 300,\ 400$ GeV
for the case $\sqrt{s}=500$ GeV and $\cot\beta=1$. 
We find that for this kinematical region the production cross section for
$A^0$ increases as $m_A$ gets larger, which looks somewhat unexpected at first glance.
%%%%%%%%%%%%%%%%%%%%%
%%%%%%%%%%%%%%%%%%%
We  examine this behaviour in more detail by computing the mass dependence of the differential
cross section. 
We plot in Fig.\ref{mass-diff-dependence}
the dependence of the differential cross section $d\sigma/dQ^2$ 
on the $A^0$ boson mass with 
$Q^2=(80)^2$, $(90)^2$ and $(100)^2$ GeV$^2$ for the case $\sqrt{s}=500$ GeV and $\cot\beta=1$. 
We see that, in the region $m_A <2m_t$, the differential cross section $d\sigma/dQ^2$ with fixed $Q^2$ increases along with $m_A$. When $m_A$ goes beyond $2m_t$, it turns to decrease. 
We observe the strong kink structure corresponding to the threshold 
effect at $m_A=2m_t\approx 346$ GeV (see Eqs.(\ref{ftau-1}) and (\ref{ftau-2})).

%%%%%%%%%%%%%%%%%%%%%
%%%%%%%%%%%%%%%%%%%

%%%%%%%%%%%%%%%%%%%%%%%%%%%%%%%%%%%%%
\section{Summary and Discussion}
%%%%%%%%%%%%%%%%%%%%%%%%%%%%%%%%%%%%%%
In this paper we have investigated the 
production
of the CP-odd Higgs boson $A^0$ which appears in the 
type-II 2HDM and the MSSM 
through $e\gamma$ collisions. In contrast to the SM Higgs boson 
$H_{SM}$ or 
the CP-even Higgs boson $h^0$ and $H^0$, the $A^0$ boson does not couple to 
$W^+W^-$ and $ZZ$ pairs because of the CP-odd nature. 
Hence $W$-boson and $Z$-boson loop 
diagrams do not contribute to the $A^0$ production at one-loop level.

The $A^0$ production 
arises via $\gamma^*\gamma$ fusion or via $Z^*\gamma$ fusion processes.
It has turned out that because of the smallness of the $e$-$e$-$Z$ and 
$Z$-$t$-$t$ couplings as well as the $Z$ boson propagator, the contribution 
from the $\gamma^*\gamma$ fusion diagrams is far more dominant over that from 
$Z^*\gamma$ fusion. Thus, in effect,  we  have to consider only the photon-exchange diagrams, 
and it makes sense to introduce the transition form 
factor of the $A^0$ boson.

Up to the electroweak one-loop order,  the top quark  triangle
diagrams are only relevant for the production of the $A^0$ boson 
when $A^0$ boson is rather light and $\tan\beta$ is not large. 
There is no scalar top-quark (stop) contribution.
Thus the production
amplitude as well as the transition form factor show much simpler 
structure compared with those of the SM Higgs boson or the CP-even Higgs
bosons. 

When the mass of the $A^0$ boson, $m_A$ is smaller than $2m_t$ the transition
form factor is a real function of $Q^2$, while if $m_A$ is larger than $2m_t$,
the transition form factor becomes complex. The production cross section of the
$A^0$ boson is given by the absolute square of the transition form
factor together with some kinematical factors.

For a fixed value of $m_A$, the differential production cross section shows
a decreasing function of $Q^2$. On the other hand, if we fix $Q^2$ and vary
the mass of $A^0$, it increases as $m_A$ for $m_A<2m_t$ and decreases for
$m_A>2m_t$. This feature is common with the total cross section.

%%%%%%%%%%%%%%%%%%%%%%%%%%%%%%%%%
%---------------- Acknowledgments  -------------------%
%\section*{Acknowledgements}
%---------------- End of Acknowledgments  ------------%
%%%%%%%%%%%%%%%%%%%%%%%%%%%%%%%%%%%%%%%%%%%%%%%%


\begin{thebibliography}{99}

\bibitem{HiggsLHC}
ATLAS Collaboration, {\sl Phys.\ Lett.}  {\bf B716}, 1 (2012); 
CMS Collaboration, {\sl Phys.\ Lett.}  {\bf B716}, 30  (2012).

\bibitem{SpinParity} 
ATLAS Collaboration, {\sl Phys.\ Lett.}  {\bf B726}, 88  (2013); 
{\sl Phys.\ Lett.} {\bf B726}, 120  (2013); 
CMS Collaboration, {\sl Phys.\ Rev.\ Lett.}  {\bf 110}, 081803  (2013).

\bibitem{ILC} 
http://www.linearcollider.org.

\bibitem{DeRoeck}
A.~De Roeck,
\lq\lq Physics at a $\gamma\gamma$, $e\gamma$ and $e^-e^-$ Option for a Linear Collider\rq\rq,
arXiv:hep-ph/0311138 (2003). 

\bibitem{Telnov:1999vz} 
  V.~I.~Telnov,
  %``High-energy photon colliders,''
  {\sl Nucl.\ Instrum.\ Meth.}  {\bf A455}, 63 (2000)
  [hep-ex/0001029]; 
 B.~Badelek {\it et al.},
%  [ECFA/DESY Photon Collider Working Group Collaboration],
%``TESLA: The Superconducting electron positron linear collider with an integrated X-ray laser 
% laboratory. Technical design report. Part 6. Appendices. Chapter 1. Photon collider at TESLA,''
  {\sl Int.\ J.\ Mod.\ Phys.}  {\bf A19}, 5097 (2004)
  [hep-ex/0108012]; 
 M.~M.~Velasco {\it et al.},
  %``Photon photon and electron photon colliders with energies below a TeV,''
  {\sl eConf}  {\bf C010630}, E3005 (2001)
  [hep-ex/0111055].

\bibitem{Melles:1999xd} 
  M.~Melles, W.~J.~Stirling and V.~A.~Khoze,
  %``Higgs boson production at the Compton collider,''
  {\sl Phys.\ Rev.}  {\bf D61}, 054015 (2000); 
 M.~Melles,
  %``Higgs physics at a gamma gamma collider,''
  {\sl Nucl.\ Phys.\ Proc.\ Suppl.} {\bf 82}, 379 (2000); 
 G.~Jikia and S.~Soldner-Rembold,
  %``Light Higgs production at the Compton collider,''
  {\sl Nucl.\ Phys.\ Proc.\ Suppl.}  {\bf 82}, 373 (2000); 
%\bibitem{MKSZ}
M.~M.~M{\"u}hlleitner, M.~Kr{\"a}mer, M.~Spira and P.~M.~Zerwas,
Phys.~Lett.~{\bf B508}~(2001)~311; 
M.~M.~M{\"u}hlleitner,
{\sl Acta Phys.\ Polon.} {\bf B37} 1127 (2006),  [hep-ph/0512232];
D.~M.~Asner, J.~B.~Gronberg, and J.~F.~Gunion, 
%\lq\lq Detecting and Studying Higgs Bosons at a Photon-Photon Collider\rq\rq,
{\sl Phys. Rev.} {\bf D67}, 035009 (2003);
 S.~J.~Brodsky,
  %``High energy photon-photon collisions at a linear collider,''
  {\sl Int.\ J.\ Mod.\ Phys.}  {\bf A20}, 7306 (2005); 
 P.~Niezurawski,
%  %``Final results for the SM Higgs-boson production at the photon collider,''
  {\sl eConf}  {\bf C050318}, 0503 (2005). 

\bibitem{BogaczETAL}
S.~A.~Bogacz et al.,
\lq\lq SAPPHiRE: a Small $\gamma\gamma$ Higgs Factory\rq\rq,
arXiv:1208.2827 [physics.acc-ph] (2012). 

\bibitem{GK}
I.~F.~Ginzburg and M.~Krawczyk,
\lq\lq Testing Higgs Physics at the Photon Collider\rq\rq,
arXiv:1310.5881 [hep-ph] (2013); 
I.~F.~Ginzburg and M.~V.~Vychugin, {\sl Physics of Atomic Nuclei}, {\bf 67}, (2004) 281.
  
\bibitem{KWSUPL}
N.~Watanabe, Y.~Kurihara, K.~Sasaki and T.~Uematsu, {\sl Phys.\ Lett.}  {\bf B728}, 202 (2014); 
  PoS (RADCOR 2013) 050; PoS (RADCOR 2013) 053;
PoS (QFTHEP 2013) 040.

\bibitem{WKUSPRD}
N.~Watanabe, Y.~Kurihara, T.~Uematsu,and K.~Sasaki, {\sl Phys. Rev.} {\bf D90}, 033015 (2014).

\bibitem{Hunter}
J.~F.~Gunion, H.~E.~Haber, G.~Kane and S.~Dawason,
\lq\lq The Higgs Hunter's Guide\rq\rq 
(Addison-Wesley, 1990).

\bibitem{GunionHaber}
J.~F.~Gunion and H.~E.~Haber,  {\sl Nucl.~Phys.}  {\bf B272}, 1  (1986).

\bibitem{ATLASchargino} 
ATLAS Collab., ATLAS-CONF-2017-039 (2017).

\bibitem{CMSchargino} 
CMS Collab., arXiv:1709.05406 (2017).

\bibitem{RPPexperiment} 
C. Patrignani {\it et al}. (Particle Data Group), Chin. Phys. C, 40, 100001 (2016) and 2017 update. 
113. Supersymmetry, Part II (Experiment).

\bibitem{PassarinoVeltman}
G.~Passarino and M.~Veltman, {\sl Nucl.~Phys.}  {\bf B160}, 151  (1979);  
G.~'t Hooft and M.J.G.~Veltman, {\sl Nucl.~Phys.} {\bf B153}, 365 (1979); 
G.J.~van Oldenborgh and J.A.M.~Vermaseren, {\sl Z. Physik}  {\bf C46} 425 (1990).

\bibitem{RomaoAndringa1997}
J.C.~Romao and S.~Andringa, {\sl Eur.~Phys. J. C.} {\bf 7}, 631 (1997).

\bibitem{H2gammas}
J.~Ellis, M.~K.~Gaillard and D.~V.~Nanopoulos, {\sl Nucl.\ Phys.}  {\bf B106}, 292 (1976); 
B.~L.~Ioffe and V.~A.~Khoze, {\sl Sov.\ J.\ Part.\ Nucl.} {\bf 9}, 50 (1978); 
M.~A.~Shifman, A.~I.~Vainshtein, M.~B.~Voloshin and V.~I.~Zakharov, {\sl Sov.\ J.\ Nucl.\ Phys.} 
 {\bf 30}, 711 (1979); {\sl Phys.\  ReV.} {\bf D85}, 013015 (2012); 
T.~G.~Rizzo, {\sl Phys.\  ReV.} {\bf D22}, (1980) 178; 
M.~B.~Gavela, G.~Girardi, C.~Malleville and P.~Sorba, {\sl Nucl.\ Phys.}  {\bf B193},257 (1981); 
W.~J.~Marciano, C.~Zhang and S.~Willenbrock, {\sl Phys.\  ReV.} {\bf D85}, 013002 (2012).

\bibitem{hMSSM}
A.~Djouadi et al., {\sl Eur.~Phys. J. C.} {\bf 73}, 2650 (2013); 
E.~Bagnaschi et al., LHCHXSWG-2015-002, CERN (2015).

\bibitem{ATLAStanbeta}
ATLAS Collab., {\sl JHEP}  {\bf 01}, 055  (2018).


\end{thebibliography}
\end{document}